\documentclass[tightenlines,preprintnumbers]{revtex4}
\begin{document}
\renewcommand\thesection {\arabic{section}}
\def\theequation{\thesection.\arabic{equation}}
\def\theref{\thesection.\arabic{equation}}
\makeatletter
\@addtoreset{equation}{section}
\makeatother
\preprint{hep-ph/0211048}
\title{On relation between unitary gauge and gauge given by 
$\xi$-limiting process}
\author{T. Kiyan\footnote{E-mail : {kyan@sci.kumamoto-u.ac.jp}}, T. Maekawa, M. Masuda and H. Taira}
\affiliation{Department of Physics, Kumamoto University, Kumamoto 860-8555}
\begin{abstract}
It is shown that in general the gauge given by the limit of $\xi \rightarrow 0$ 
in $R_\xi$ gauge does not necessarily agree with the unitary gauge by examining 
the symmetry breaking of two steps from ${\rm SU(3) \otimes U(1)}_N$ to 
${\rm SU(2) \otimes U(1)}_Y$ and then to ${\rm U(1)}_{em}$.

\vspace{0.2cm}
\end{abstract}
\maketitle
%
\section{Introduction}

It is usually stated that the special gauge given by $\xi \rightarrow 0$ in 
$R_\xi$ gauge\cite{R-gauge,FP-ghost} is the unitary (U) gauge\cite{SW,AL}, 
in this gauge only the exchanges of physical vector and physical 
Higgs particles\cite{PWH} may be treated because the ghost fields 
and the Goldstone bosons\cite{GOLD} drop out, and the U gauge is renormalizable 
if ultraviolet divergences resulting from the vector propagators are handled carefully. 
The U gauge is defined by the condition that the scalar field $\phi(x)$ 
after the gauge fixing has no components in the subspace spanned by the Goldstone bosons, 
which is the space spanned by $L^iv$ with the real representation matrices 
$L^i$ corresponding to the symmetry breaking generators and with 
the vacuum expectation value $v$ of a scalar field. On the other hand, 
the $R_\xi$ gauge is usually defined by the condition that the gauge fixing 
is given through the gauge fields together with the ghost fields\cite{FP-ghost}. 
Though their definition is different from each other, 
the gauge given by $\xi\rightarrow 0$ in $R_\xi$ agrees with the U gauge 
in many cases such as in the standard model(SM). It is, however, 
expected that the special gauge and the U gauge will not necessarily agree 
because the $R_\xi$ gauge is chosen with no transition terms resulting from 
the covariant kinetic energy term after the symmetry breaking 
while these transition terms do not vanish in the U gauge in general\cite{TTMH}. 

In this note, we study the model with the gauge group ${\rm SU(3) \otimes U(1)}_N$ 
without the color symmetry and introduce the scalar fields, $\chi, \rho, \eta$, 
in order to make our discussion clear as well as the gauge fields. 
It is shown that when the spontaneous symmetry breaking (SSB) occurs through 
two steps such as ${\rm SU(3) \otimes U(1)}_N \rightarrow {\rm SU(2) \otimes U(1)}_Y 
\rightarrow {\rm U(1)}_{em}$ the condition of the unitary gauge is not satisfied for 
the scalar particles remaining without vanishing in the limit of $\xi \rightarrow 0$.
\section{Preliminaries}

Let us introduce the scalar fields which may be irreducible components of a scalar field 
under the group ${\rm SU(3) \otimes U(1)}_N$ as follows
\begin{eqnarray}
&&
\chi = 
\left(
  \begin{array}{c}
	\chi^- \\
	\chi^{--} \\
	\chi^0 
  \end{array}
\right) \sim (3,-1),\nonumber\\
&&
\rho = 
\left(
  \begin{array}{c}
	\rho^+ \\
	\rho^{0} \\
	\rho^{++} 
  \end{array}
\right) \sim (3,1),\label{2.1}\\
&&
\eta = 
\left(
  \begin{array}{c}
	\eta^0 \\
	\eta^{-} \\
	\eta^+ 
  \end{array}
\right) \sim (3,0)\nonumber,
\end{eqnarray}
where the charge operator is given by $Q=T^3 -{\sqrt 3}T^8 +N$ with the generators 
$T^j, N$ of ${\rm SU(3)}$ and ${\rm U(1)}_N$. The renormalizable gauge invariant Lagrangian 
for the scalar and gauge fields is given by
\begin{eqnarray}
{\cal L} = -\frac{1}{4}\sum_a V^a_{\mu\nu}V^{a\mu\nu}
           +\sum_\phi({\cal D}^\mu\phi)^\dag({\cal D}\phi)-V(\phi),\label{2.2}
\end{eqnarray}
where $\phi$ stands for $\chi, \rho$ and $\eta$, and the covariant derivative for 
$\phi$ is defined by
\begin{eqnarray*}
{\cal D}_\mu\phi = 
\Bigl(\partial_\mu-i\frac{g}{2}\sum_{j=1}^8\lambda^jA^j_\mu-ig_Ny_\phi B_\mu\Bigr)\phi,
\end{eqnarray*}
with $y_\phi =-1,1$ and 0 for $\phi=\chi, \rho$ and $\eta$. The field strengths of 
the gauge fields are as follows
\begin{eqnarray}
V^a_{\mu\nu} = \partial_\mu V^a_\nu
              -\partial_\nu V^a_\mu
              +g_a\sum_{bc}^9f_{abc}V^b_\mu V^c_\nu,\label{2.3}
\end{eqnarray}
with $V^a_\mu=A^a_\mu, g_a=g (a=1,\cdots, 8), V^9_\mu=B_\mu, g_9=g_N$ and 
the structure constants of the gauge group ${\rm SU(3) \otimes U(1)}_N$ with $f_{ab9}=0$. 
In what follows the summation convention will be used unless stated otherwise.

The potential of the scalar fields  is given as follows
\begin{eqnarray}
V(\phi)
  &=&c_1\bigl(\chi^\dag\chi-\frac{1}{2}\chi_v^2\bigr)^2
    +c_2\bigl(\rho^\dag\rho-\frac{1}{2}\rho_v^2\bigr)^2
    +c_3\bigl(\eta^\dag\eta-\frac{1}{2}\eta_v^2\bigr)^2\nonumber\\
  &+&c_4\bigl(\chi^\dag\chi-\frac{1}{2}\chi_v^2\bigr)\bigl(\rho^\dag\rho-\frac{1}{2}\rho_v^2\bigr)
    +c_5\bigl(\chi^\dag\chi-\frac{1}{2}\chi_v^2\bigr)\bigl(\eta^\dag\eta-\frac{1}{2}\eta_v^2\bigr)\nonumber\\
  &+&c_6\bigl(\rho^\dag\rho-\frac{1}{2}\rho_v^2\bigr)\bigl(\eta^\dag\eta-\frac{1}{2}\eta_v^2\bigr)\nonumber\\
  &+&c_7\bigl(\chi^\dag\rho\bigr)\bigl(\chi^\dag\rho\bigr)^{\dag}
    +c_8\bigl(\chi^\dag\eta\bigr)\bigl(\chi^\dag\eta\bigr)^{\dag}
    +c_9\bigl(\rho^\dag\eta\bigr)\bigl(\rho^\dag\eta\bigr)^{\dag},\label{2.4}
\end{eqnarray}
where the $c_a, \chi_v, \rho_v$ and $\eta_v$ are some real constants and (\ref{2.4}) 
is written by assuming a symmetry under $\phi \rightarrow -\phi$. 

The scalar fields may be treated with the hermitian fields by writing 
$\phi_j=(\phi_{2j}+i\phi_{1j})/\sqrt{2}$ with the hermitian fields $\phi_{1j}$ and 
$\phi_{2j}$, and the notation $\hat\phi$ with the components 
${\hat\phi}_{\alpha j}=\phi_{\alpha j}$ ($\alpha =1,2$) is used. 
Then, the covariant derivative corresponding to that below (\ref{2.2}) can be 
rewritten as follows
\begin{eqnarray}
{\hat{\cal D}_\mu{\hat\phi}}
\equiv\bigl(\partial_\mu-i({\hat L}^a_-+{\hat L}^a_+)V^a_\mu\bigr){\hat\phi},\label{2.5}
\end{eqnarray}
where the representation matrices, $\hat L^a_\pm$, for $\hat\phi$ are given by
\begin{eqnarray}
&&
{\hat L}^a_- = I_2\otimes{L}^a_-,\ \ 
{\hat L}^a_+ = \tau_2\otimes{L}^a_+,\label{2.6}\\
&&
L^a_\pm = \frac{1}{2}(L^a\pm L^{aT}),\ \ 
L^a = \frac{1}{2}\lambda^a\ \ (a=1,2,\cdots, 8),\ \ 
L^9 = y_\phi I^3.\nonumber
\end{eqnarray}
It is noted that the representation matrices ${\hat L}^a_{\pm}$ for $\hat\phi$ 
are hermitian and antisymmetric and thus they are equivalent to some of 
the representation matrices of the orthogonal group ${\rm SO(6)}$. 
The action of the ${\hat L}^a_\pm$ on ${\hat\phi}$ is given by 
$({\hat L}^a_\pm{\hat\phi})_{\alpha j}
=(I_2\ {\rm or}\ \tau_2)_{\alpha\beta}L^a_{\pm jk}{\hat\phi}_{\beta k}$ with 
$(I_2)_{\alpha\beta}=\delta_{\alpha\beta}$ and $(\tau_2)_{\alpha\beta}
=-i\epsilon_{\alpha\beta}=i\epsilon_{\beta\alpha},\ \epsilon_{12}=1,$ and 
the following relations hold
\begin{eqnarray}
&&
\psi^\dag\phi = \frac{1}{2}{\hat\psi}^T(I_2+\tau_2)\otimes I^3{\hat\phi},\nonumber\\
&&
({\cal D}^\mu\psi)^\dag{\cal D}_\mu\phi = 
\frac{1}{2}(\hat{\cal D}^\mu\hat\psi)^T
(I_2+\tau_2)\otimes I^3\hat{\cal D}_\mu{\hat\phi}.\label{2.7}
\end{eqnarray}

We use the notation $\Phi=(\chi,\ \rho,\ \eta)$ when the scalar fields 
are considered as the irreducible components of a scalar $\Phi$. 
Then the covariant derivative for $\Phi$ can be written as follows
\begin{eqnarray}
&&
{\cal D}_\mu\Phi = (\partial_\mu-iK^\Phi_\mu)\Phi,\label{2.8}\\
&&
K^\Phi_\mu = K^\chi_\mu\oplus K^\rho_\mu\oplus K^\eta_\mu,\nonumber
\end{eqnarray}
where $\oplus$ means a direct sum of the matrices and the notation $K^\phi_\mu$ 
denotes the quantity $g\lambda\cdot A_\mu /2+g_Ny_\phi B_\mu$ given below (\ref{2.2}). 
The covariant derivative for $\hat\Phi$ corresponding to (\ref{2.8}) is given as follows
\begin{eqnarray}
&&
{\hat{\cal D}}_\mu\hat\Phi = (\partial_\mu-iK^{\hat\Phi}_\mu)\hat\Phi,\label{2.9}\\
&&
K^{\hat\Phi}_\mu = 
     I_2\otimes (K^\chi_{\mu -}\oplus K^\rho_{\mu -}\oplus K^\eta_{\mu -})
    +\tau_2\otimes (K^\chi_{\mu +}\oplus K^\rho_{\mu +}\oplus K^\eta_{\mu +}),\nonumber\\
&&
K^\phi_{\mu\pm} = \frac{1}{2}(K^\phi_\mu\pm K^{\phi T}_\mu).\nonumber
\end{eqnarray}

The covariant kinetic energy for the scalar in (\ref{2.2}) can be rewritten as follows
\begin{eqnarray}
\sum_\phi({\cal D^\mu}\phi)^\dag({\cal D}_\mu\phi)
  &=&({\cal D}^\mu\Phi)^\dag{\cal D}_\mu\Phi\nonumber\\
  &=&\frac{1}{2}({\hat{\cal D}}^\mu{\hat\Phi})^T 
  (I_2+\tau_2)\otimes (I^3\oplus I^3\oplus I^3){\hat{\cal D}}_\mu{\hat\Phi}.\label{2.10}
\end{eqnarray}
%
%
\section{SSB and kinetic energy term of scalars}

By introducing the VEV($\chi_v$) for $\chi$, the symmetry ${\rm SU(3) \otimes U(1)}_N$ 
is broken down to ${\rm SU(2) \otimes U(1)}_Y$ with the hypercharge $Y=N-\sqrt{3}T^8$. 
The field is rewritten as follows
\begin{eqnarray}
\chi = \frac{1}{\sqrt 2}(\chi_v+\chi^\prime) 
       = \frac{1}{\sqrt 2}\left(
           \begin{array}{c}
             \chi_1^- \\
             \chi_2^{--} \\
             \chi_v+\chi_3^0 
           \end{array} \right).\label{3.1}
\end{eqnarray}
The hermitian field $\hat\chi$ corresponding to $\chi$ is written as follows
\begin{eqnarray*}
  {\hat\chi} = {\hat\chi}_v+{\tilde\chi},
\end{eqnarray*}
with $(\hat\chi_v)_{\alpha j}=\chi_v\delta_{\alpha 2}\delta_{j3}$.

By fixing the gauge through the Higgs mechanism, the $\chi$ changes to the form
\begin{eqnarray}
\chi = \frac{1}{\sqrt 2}\left(
           \begin{array}{c}
             0 \\
             0 \\
             \chi_v+ \chi_h
           \end{array}\right),\ \ \ 
         \chi_h^\dag = \chi_h,\label{3.2}
\end{eqnarray}
where the same notation for the $\chi$ before and after a transformation is used because 
the confusion will not occur and in what follows the same will be made unless stated otherwise. 
By the unitary transformation to fix the gauge, the fields $\rho$ and $\eta$ will be affected 
but have the form  corresponding to those before the symmetry breaking.

It is easily seen that the condition of the unitary gauge\cite{SW,AL} is satisfied 
in the case of the breaking because the following relations hold for the matrices 
corresponding to the broken generators
\begin{eqnarray}
&&
\hat\chi^T_v I_2 \otimes \lambda^a {\tilde\chi}_h =0, \ \ (a=5,7),\nonumber\\
&&
\hat\chi^T_v \tau_2 \otimes \lambda^a {\tilde\chi}_h =0, \ \ (a=4,6),\nonumber\\
&&
\hat\chi^T_v \tau_2 \otimes (-\sqrt {3}\lambda^8/2+1) {\tilde\chi}_h =0,\label{3.3}
\end{eqnarray}
with $(\tilde\chi_h)_{\alpha j}=\chi_h\delta_{\alpha 2}\delta_{j3}$.

Next, by the VEV of $\rho$, the symmetry is broken down from ${\rm SU(2) \otimes U(1)_Y}$ 
to ${\rm U(1)_{em}}$ and the $\rho$ may be written as follows
\begin{eqnarray}
\rho = \frac{1}{\sqrt 2}(\rho_v+\rho^\prime)
     = \frac{1}{\sqrt 2}\left(
       \begin{array}{c}
        \rho_1^+ \\
        \rho_v+\rho_2^0 \\
        \rho_3^{++}
       \end{array}\right).\label{3.4}
\end{eqnarray}
The hermitian field $\hat \rho$ is written  as follows
\begin{eqnarray*}
\hat\rho = \hat\rho_v+\tilde\rho.
\end{eqnarray*}

By fixing the gauge through the Higgs mechanism as in for the $\chi$, 
the $\rho$ may be written in the form
\begin{eqnarray}
\rho = \frac{1}{\sqrt 2}\left(
           \begin{array}{c}
            0 \\
            \rho_v+\rho_h \\
            \rho^{++}_{3h}
           \end{array} \right),\ \ \ 
          \rho_h^\dag = \rho_h.\label{3.5}
\end{eqnarray}
Then, the condition of the unitary gauge is satisfied as in the case of the SM, i.e.,
\begin{eqnarray}
&&
\hat\rho^{T}_{v} I_{2}\otimes\lambda^{2}{\tilde\rho}_h = 0,\nonumber\\
&&
\hat\rho^{T}_{v} \tau_{2}\otimes\lambda^{1}{\tilde \rho}_h = 0,\nonumber\\
&&
\hat\rho^{T}_{v} \tau_{2}\otimes\bigl((\lambda^{3}+\sqrt{3}\lambda^{8})/2-1\bigr){\tilde\rho}_h = 0,\label{3.6}
\end{eqnarray}
with the nonzero components $\{\hat\rho_v\}_{\alpha j} = \rho_v\delta_{\alpha 2}\delta_{j2},\ 
{\tilde\rho}_{\alpha 2} = \rho_{h}\delta_{\alpha 2}$ and ${\tilde\rho}_{\alpha 3}$ and thus 
the only symmetry of the ${\rm U(1)_{em}}$ remains with the charge operator $Q(=T^3+Y)$. 
It is noted that the third component of the $\chi$ is not affected 
but that of $\eta$ is affected by a transformation of the gauge fixing at this stage.

Similarly, the VEV of the $\eta$ is introduced as follows
\begin{eqnarray}
\eta = \frac{1}{\sqrt 2}(\eta_v+\eta^\prime)
       = \frac{1}{\sqrt 2}\left(
           \begin{array}{c}
            \eta_v+\eta_1^0 \\
            \eta_2^{-} \\
            \eta_3^+ 
           \end{array} \right).\label{3.7}
\end{eqnarray}
It is evident that the VEV of the $\eta$ is independent of the symmetry breaking because 
the relation $\tau_2 \otimes Q{\hat \eta}_v=0$ holds and there remains no symmetry breaking generator.

It is easily seen that the above conditions (\ref{3.3}) and (\ref{3.6}) do not change even in the case 
by one step of breaking from ${\rm SU(3) \otimes U(1)}_N$ to ${\rm U(1)}_{em}$. 
Explicitly, with a well known method one parametrizes the $\chi$ in the form 
$U^\dag_\chi (\chi_v+\chi_h)/\sqrt 2$ with the $\chi$ in (\ref{3.2}) and the $\rho$ 
in the form $U^\dag_\chi U^\dag_\rho (\rho_v+\rho_h)/\sqrt 2$ with the $\rho$ of (\ref{3.5}), 
and then the unitary transformation by $U_\rho U_\chi$  may be carried out to give the desired result.

The covariant kinetic energy terms of the scalar fields in terms of the tilde fields are rewritten as follows
\begin{eqnarray}
\sum_\phi ({\cal D}^\mu \phi)^\dag {\cal D}_\mu\phi 
     &=&\frac{1}{2}\sum_\phi\left [(\hat {\cal D}^\mu\tilde\phi)^T \hat{\cal D}_\mu\tilde\phi\right.\nonumber\\
     &+&\left.\Bigl\{\hat v^T(I_2+\tau_2) \otimes K^{\phi\mu}K^\phi_\mu\tilde\phi
       +\tilde\phi^T(I_2+\tau_2) \otimes K^{\phi\mu}K^\phi_\mu\hat v \Bigr\}\right.\nonumber\\
     &+&\left. 2i\hat v^TK^\phi_\mu\partial^\mu\tilde\phi
       +\hat v^TK^{\phi\mu}K^{\phi}_{\mu}\hat v\right],\label{3.8}
\end{eqnarray}
where $\hat v$ stands for $\hat\chi_v,\ \hat \rho_v,\ \hat \eta_v$. 
The last term in (\ref{3.8}) gives the mass to the gauge bosons and rewritten in terms of 
the mass eigenstates as follows
\begin{eqnarray}
M^2_WW^{\mu +}W^-_\mu +M^2_XX^{\mu +}X^-_\mu +M^2_YY^{\mu ++}Y^{--}_\mu
+\frac{1}{2}\left(M^2_{Z_1}Z^\mu_1Z_{1\mu}+M^2_{Z_2}Z^\mu_2Z_{2\mu}\right),\label{3.9}
\end{eqnarray}
where
\begin{eqnarray*}
&&
\sqrt 2 W^{\pm}_\mu = A^1_\mu\mp iA^2_\mu,\ \ \ 
\sqrt 2 X^{\pm}_\mu  = A^4_\mu\pm iA^5_\mu,\ \ \ 
\sqrt 2 Y^{\pm \pm}_\mu = A^6_\mu\pm iA^7_\mu ,\\
&&
A_\mu = s_WA^3_\mu+c_W\Bigl(-\sqrt 3t_WA^8_\mu+\sqrt {1-3t^2_W}B_\mu\Bigr),\\
&&
Z_\mu = c_WA^3_\mu -s_W\Bigl(-\sqrt 3t_WA^8_\mu+\sqrt {1-3t^2_W}B_\mu\Bigr),\ \ \ 
Z^\prime_\mu = \sqrt {1-3t^2_W}A^8_\mu+\sqrt 3t_W B_\mu,\\
&&
Z_{1\mu} = \cos\phi Z_\mu+\sin\phi Z^\prime_\mu,\ \ \ 
Z_{2\mu} = -\sin\phi Z_\mu+\cos\phi Z^\prime_\mu ,\\
&&
M^2_W = \frac{g^2}{4}\Bigl(\rho^2_v+\eta^2_v\Bigr),\ \ \ 
M^2_X = \frac{g^2}{4}\Bigl(\chi^2_v+\eta^2_v\Bigr),\ \ 
M^2_Y = \frac{g^2}{4}\Bigl(\chi^2_v+\rho^2_v\Bigr),\\
&&
M^2_{Z_1} = 
\frac{1}{2}\left[M^2_Z+M^2_{Z^\prime}-\sqrt {(M^2_{Z^\prime}-M^2_Z)^2+4(M_{ZZ^\prime})^4}\ \right],\\
&&
M^2_{Z_2} = 
\frac{1}{2}\left[M^2_Z+M^2_{Z^\prime}+\sqrt {(M^2_{Z^\prime}-M^2_Z)^2+4(M_{ZZ^\prime})^4}\ \right],\\
&&
c_W = \cos\theta_W,\ \ \ 
t_W = \tan\theta_W = \frac{g_N}{\sqrt {g^2+3g^2_N}}, \ \ \ 
\tan\phi = \frac{M^2_Z -M^2_{Z_1}}{M^2_{ZZ^\prime}},\ \ \  
M^2_Z = \frac{M^2_W}{c^2_W},\\
&&
M^2_{Z^\prime} = 
\frac{1}{3\bigl(1-3t^2_W\bigr)}\Bigl[\bigl(2+3t^2_W\bigr)M^2_Y
+\bigl(2-3t^2_W\bigr)M^2_X-\bigl(1+3t^2_W\bigr)\bigl(1-3t^2_W\bigr)M^2_W\Bigr],\\
&&
M^2_{ZZ^\prime} = \frac{1}{c_W\sqrt {3\bigl(1-3t^2_W\bigr)}}\Bigl[M^2_Y-M^2_X+3t^2_WM^2_W\Bigr].
\end{eqnarray*}

The third term on the right side in (\ref{3.8}) remains without vanishing in the Higgs mechanism 
in contrast to the cases of the SM and the $R_\xi$ gauge and is given as follows
\begin{eqnarray}
&&
ig\rho_v\bigl(Y^{--}_\mu\partial^\mu\rho^{++}_h-Y^{++}_\mu\partial^\mu\rho^{--}_h\bigr)
+ig\eta_v\bigl(W^{+}_\mu\partial^\mu\eta^{-}_h-W^{-}_\mu\partial^\mu\eta^{-\dag}_h\bigr)\nonumber\\
&&
+ig\eta_v\bigl(X^{-}_\mu\partial^\mu\eta^{+}_h-X^{+}_\mu\partial^\mu\eta^{+\dag}_h\bigr)
-g\eta_v\bigl(A^3_\mu+\frac{1}{\sqrt 3}A^8_\mu\bigr)\partial^\mu\tilde\eta_{11},\label{3.10}
\end{eqnarray}
where $\eta^{\pm}_{h}=\eta^{\pm}$. 
The last gauge term in (\ref{3.10}) can be expressed in terms of the $Z_\mu$ and $Z^\prime_\mu$ as follows
\begin{eqnarray*}
A^3_\mu+\frac{1}{\sqrt 3}A^8_\mu = \frac{1}{c_W}Z_\mu+\frac{1}{\sqrt 3}\sqrt {1-3t^2_W}Z^\prime_\mu ,
\end{eqnarray*}
which means that the transition between $A_\mu$ and $\eta_{11}$ does not occur as is desired. 
Even with use of the $R_\xi$ gauge in the stage of the symmetry breaking of ${\rm SU(2) \otimes U(1)}_Y$ 
to ${\rm U(1)}_{em}$ these terms remain without vanishing. The existence of these transition terms gives 
an effect on the propagator through the mixing of the scalar and gauge fields, and then gives rise to 
disagreement with the limit of $\xi \rightarrow 0$ in the $R_\xi$ gauge in which these transition terms 
disappear from the Lagrangian by the gauge fixing term.
%
%
\section{Gauge fixing and ghost terms in $R_\xi$ gauge}

In this gauge, the scalar fields have three non-zero components as in (\ref{3.1}) and (\ref{3.4}). 
It is convenient to use the hermitian fields such as that below (\ref{3.1}). 
Then, the Lagrangian is given by these fields and a brief explanation will be given\cite{TTMH}.

The second order terms on the fields in the Lagrangian are related with the propagators and 
one will give these terms below. The kinetic energy terms of the scalar fields are written as follows
\begin{eqnarray}
\sum_\phi (\partial^\mu\phi)^\dag\partial_\mu\phi 
    &=& \frac{1}{2}\biggl[\frac{1}{\rho^2_v+\eta^2_v}
           \Bigl\{(\rho_v \partial^\mu{\tilde\eta}_{12}-\eta_v \partial^\mu{\tilde\rho}_{11})^2
           +(\eta_v \partial^\mu{\tilde\eta}_{12}+\rho_v \partial^\mu{\tilde\rho}_{11})^2\Bigr\}\nonumber\\
    &+& \frac{1}{\rho^2_v+\eta^2_v}
           \Bigl\{(\rho_v \partial^\mu{\tilde\eta}_{22}+\eta_v \partial^\mu{\tilde\rho}_{21})^2
           +(\eta_v \partial^\mu{\tilde\eta}_{22}-\rho_v \partial^\mu{\tilde\rho}_{21})^2\Bigr\}\nonumber\\
    &+& \frac{1}{\chi^2_v+\rho^2_v}
           \Bigl\{(\chi_v \partial^\mu{\tilde\rho}_{13}-\rho_v \partial^\mu{\tilde\chi}_{12})^2
           +(\chi_v \partial^\mu{\tilde\chi}_{12}+\rho_v \partial^\mu{\tilde\rho}_{13})^2\Bigr\}\nonumber\\
    &+& \frac{1}{\chi^2_v+\rho^2_v}
           \Bigl\{(\chi_v \partial^\mu{\tilde\rho}_{23}+\rho_v \partial^\mu{\tilde\chi}_{22})^2
           +(\chi_v \partial^\mu{\tilde\chi}_{22}-\rho_v \partial^\mu{\tilde\rho}_{23})^2\Bigr\}\nonumber\\
    &+& \frac{1}{\chi^2_v+\eta^2_v}
           \Bigl\{(\chi_v \partial^\mu{\tilde\eta}_{23}+\eta_v \partial^\mu{\tilde\chi}_{21})^2
           +(\chi_v \partial^\mu{\tilde\chi}_{21}-\eta_v \partial^\mu{\tilde\eta}_{23})^2\Bigr\}\nonumber\\
    &+& \frac{1}{\chi^2_v+\eta^2_v}
           \Bigl\{(\chi_v \partial^\mu{\tilde\eta}_{13}-\eta_v \partial^\mu{\tilde\chi}_{11})^2
           +(\chi_v \partial^\mu{\tilde\chi}_{11}+\eta_v \partial^\mu{\tilde\eta}_{13})^2\Bigr\}\biggr]\nonumber\\
    &+& (\partial^\mu\tilde\chi_{\alpha 3})^2+(\partial^\mu\tilde\rho_{\alpha 2})^2
           +(\partial^\mu\tilde\eta_{\alpha 1})^2,\label{4.1}
\end{eqnarray}
where each term is rewritten in a linear form except for the last line for latter convenience. 
Similarly, the second order terms in the potential (\ref{2.4}) are given as follows
\begin{eqnarray}
V &\sim& \frac{c_7}{4}\Bigl\{(\chi_v {\tilde\rho}_{13}-\rho_v {\tilde\chi}_{12})^2
              +(\rho_v {\tilde\chi}_{22}+\chi_v {\tilde\rho}_{23})^2\Bigr\}\nonumber\\
       &+& \frac{c_8}{4}\Bigl\{(\chi_v {\tilde\eta}_{13}-\eta_v {\tilde\chi}_{11})^2
              +(\chi_v {\tilde\eta}_{23}+\eta_v {\tilde\chi}_{21})^2\Bigr\}\nonumber\\
       &+& \frac{c_9}{4}\Bigl\{(\rho_v {\tilde\eta}_{12}-\eta_v {\tilde\rho}_{11})^2
              +(\rho_v {\tilde\eta}_{22}+\eta_v {\tilde\rho}_{21})^2\Bigr\}\nonumber\\
       &+& \frac{1}{2}\bigl(\tilde\chi_{23},\ \tilde\rho_{22},\ \tilde\eta_{21}\bigr)
              M^2_{(\chi,\rho,\eta)}\left(
               \begin{array}{c}
                \tilde\chi_{23} \\
                \tilde\rho_{22} \\
                \tilde\eta_{21} 
               \end{array} \right),\label{4.2}
\end{eqnarray}
where
\begin{eqnarray*}
M^2_{(\chi,\rho,\eta)} = \left(
    \begin{array}{ccc}
      2c_1\chi^2_v & c_4\chi_v \rho_v & c_5 \chi_v \eta_v \\
      c_4\chi_v \rho_v & 2c_2\rho^2_v & c_6\rho_v \eta_v \\
      c_5 \chi_v \eta_v & c_6\rho_v \eta_v & 2c_3\eta^2_v 
    \end{array} \right).
\end{eqnarray*}
It is noted that the second order terms of the scalar fields in the Higgs mechanism 
are given from (\ref{4.1}) and (\ref{4.2}) by putting $\tilde\chi_{\alpha j} \rightarrow 
\chi_h \delta_{\alpha 2}\delta_{j 3},\ \ \tilde\rho_{\alpha 1} \rightarrow  0,\ \ 
\tilde\rho_{\alpha 2} \rightarrow \rho_h \delta_{\alpha 2}$ with all other terms.

The gauge fixing Lagrangian is given without using auxiliary  fields by\cite{TTMH}
\begin{eqnarray}
{\cal L}_{gf} = -\frac{\xi}{2}\sum_{a=1}^9
         \Bigl(\partial^\mu V^a_\mu-i\frac{1}{\xi}g^av^TL^{\tilde\Phi a}\tilde\Phi\Bigr)^2,\label{4.3}
\end{eqnarray}
where
\begin{eqnarray*}
&&
L^{\tilde\Phi a}\equiv I_2 \otimes \bigl(L^{\chi a}_-\oplus L^{\rho a}_-\oplus L^{\eta a}_-\bigr)
                        +\tau_2 \otimes \bigl(L^{\chi a}_+\oplus L^{\rho a}_+\oplus L^{\eta a}_+\bigr),\\
&&
L^{\phi a}_\pm  = \frac{1}{2}\bigl(L^{\phi a}\pm L^{\phi a T}\bigl),\\
&&
L^{\phi  a} = L^a = \frac{1}{2}\lambda^a\ \  (a=1,2,\cdots , 8);\ \ y_\phi I^3\ \ (a=9),\\
&&
\tilde\Phi = \bigl(\tilde\chi,\ \tilde\rho,\ \tilde\eta\bigr)^T.
\end{eqnarray*}

It is noted that the cross terms from (\ref{4.3}) give the total divergence together with 
the third term in (\ref{3.8}) and thus these terms are omitted from the Lagrangian. 
The explicit expression of (\ref{4.3}) except for the cross terms is given as follows
\begin{eqnarray}
{\cal L}_{gf} 
&\sim& -\xi\left[\bigl(\partial^\mu W^+_\mu\bigr)^2
           +\bigl(\partial^\mu X^+_\mu\bigr)^2
           +\bigl(\partial^\mu Y^{++}_\mu\bigr)^2
           +\frac{1}{2}\bigl(\partial^\mu Z_{1\mu}\bigr)^2
           +\frac{1}{2}\bigl(\partial^\mu Z_{2\mu}\bigr)^2
           +\frac{1}{2}\bigl(\partial^\mu A_\mu\bigr)^2\right]\nonumber\\
     &-& \frac{g^2}{8\xi}\biggl[\bigl(\eta_v {\tilde\eta}_{12}+\rho_v {\tilde\rho_{11}}\bigr)^2
           +\bigl(\eta_v \tilde\eta_{22}-\rho_v \tilde\rho_{21}\bigr)^2
           +\bigl(\chi_v \tilde\chi_{11}+\eta_v \tilde\eta_{13}\bigr)^2
           +\bigl(\chi_v \tilde\chi_{21}-\eta_v \tilde\eta_{23}\bigr)^2\nonumber\biggr.\\
     &+& \biggl. \bigl(\chi_v \tilde\chi_{12}+\rho_v \tilde\rho_{13}\bigr)^2
           +\bigl(\chi_v \tilde\chi_{22}-\rho_v \tilde\rho_{23}\bigr)^2\biggr]
           -\frac{1}{2\xi}\biggl\{M^2_+\bigl(\tilde\rho_{M+}\bigr)^2
           +M^2_-\bigl(\tilde\eta_{M-}\bigr)^2\biggr\},\label{4.4}
\end{eqnarray}
where
\begin{eqnarray*}
&&
\tilde\chi_0 = a\bigl(\rho_v\eta_v\tilde\chi_{13}
                    +\chi_v\eta_v\tilde\rho_{12}
                    +\chi_v\rho_v\tilde\eta_{11}\bigr),\ \ \ 
\tilde\rho_{M_+} = \cos\theta\tilde\rho_- +\sin\theta\tilde\eta_+ ,\\
&&
\tilde\eta_{M_-} = -\sin\theta\tilde\rho_- +\cos\theta\tilde\eta_+,\ \ \ 
\tilde\rho_- = b\bigl(\chi_v\tilde\chi_{13}+\rho_v\tilde\rho_{12}-2\eta_v\tilde\eta_{11}\bigr),\\
&&
\tilde\eta_+ = c\Bigl\{\chi_v\bigl(\rho_v^2+2\eta^2_v\bigr)\tilde\chi_{13}
                    -\rho_v\bigl(\chi_v^2+2\eta^2_v\bigr)\tilde\rho_{12}
                    +\eta_v\bigl(\chi_v^2-\rho^2_v\bigr)\tilde \eta_{11}\Bigr\},\\
&&
a^2\bigl(\chi^2_v\rho^2_v+\chi^2_v\eta^2_v+\eta^2_v\rho^2_v\bigr) = 1,\ \ \ 
b^2\bigl(\chi^2_v+\rho^2_v+4\eta^2_v\bigr)=1,\ \ \ 
c=ab,\\
&&
\tan\theta = \frac{1}{2M^2_{23}}
                  \left[M^2_{33}-M^2_{22}
                 -\sqrt {\bigl(M^2_{33}-M^2_{22}\bigr)^2+4\bigl(M^2_{23}\bigr)^2}\ \right],\\
&&
M^2_{22} = b^2\left[\frac{g^2}{3}
               \Bigl(\chi^4_v+\rho^4_v+4\eta^4_v -\rho^2_v\chi^2_v+2\eta^2_v\chi^2_v+2\eta^2_v\rho^2_v\Bigr)
               +g^2_N\Bigl(\chi^2_v-\rho^2_v\Bigr)^2\right],\\
&&
M^2_{33} = \frac{4b^2}{a^2}\left(\frac{g^2}{4}+g^2_N\right),\ \ \ 
M^2_{23} = \frac{2b^2}{a}\biggl(\chi^2_v-\rho^2_v\biggr)\left(\frac {g^2}{4}+g^2_N\right),\\
&&
M^2_\pm = \frac{1}{2}\left[\frac{g^2}{3}\Bigl(\chi^2_v+\rho^2_v+\eta^2_v\Bigr)
               +g^2_N\Bigl(\chi^2_v+\rho^2_v\Bigr)\right.\\
&&\left.\quad\quad
\mp\sqrt{\Bigl\{\frac{g^2}{3}\Bigl(\chi^2_v+\rho^2_v+\eta^2_v\Bigr)
    +g^2_N\Bigl(\chi^2_v+\rho^2_v\Bigr)\Bigr\}^2
    -\frac{4g^2}{3a^2}\Bigl(\frac{g^2}{4}+g^2_N\Bigr)}\ \right].
\end{eqnarray*}

The Lagrangian for the ghost fields is given by
\begin{eqnarray}
{\cal L}_{gh} = 
i\left[\partial^{\mu}{\overline C^{a}}\bigl(\partial_{\mu}C^{a}+g_{a}f_{abc}V^{b}_{\mu}C^{c}\bigr)
-\frac{g_{a}g_{b}}{\xi}{\overline C^{a}}v^{T}
L^{\tilde{\Phi}a}L^{\tilde{\Phi}b}\bigl(v+\tilde\Phi\bigr)C^{b}\right].\label{4.5}
\end{eqnarray}

The masses of the ghost fields are determined from the last term in (\ref{4.5}) and given with 
the expression corresponding to (\ref{3.9}) for the vector fields except for the factor $\xi$. 
The notations such as $C^\pm_W$ and ${\overline C}^\pm_W$ are used for the ghost fields 
with the definite mass corresponding to these of the vector field such as $W^\pm_\mu$.
%
%
\section{Discussion}

In the Higgs mechanism the field $\chi$ is changed to the form of (\ref{3.2}) in the symmetry breaking of 
${\rm SU(3) \otimes U(1)}_N$ to ${\rm SU(2) \otimes U(1)}_Y$ but the $\rho,\ \eta$ have 
the three components which are affected by the gauge transformation. 
In the violation of ${\rm SU(2) \otimes U(1)}_Y$ to ${\rm U(1)}_{em}$ the $\rho$ is fixed to 
the form (\ref{3.5}) by the gauge transformation. 
Thus, in the case of the U gauge the third term of (\ref{3.8}) (or the transition terms (\ref{3.10})) 
appears without vanishing and can not be eliminated by some physical procedure because there remains 
only to fix the gauge of ${\rm U(1)}_{em}$. 
Thus, the propagators for the gauge and scalar fields are given through a combination of these fields. 
For instance, for a pair of $A^6_\mu$ and $\rho_{h13}$ their propagator satisfies the equation
\begin{eqnarray*}
\left(
 \begin{array}{cc}
 \bigl(\partial^2+M^2_Y\bigr)g^{\rho\nu}-\partial^\rho\partial^\nu & 
 -\frac{1}{2}g\rho_v\partial^\nu \\
 \frac{1}{2}g\rho_v\partial^\rho & 
 -\bigl(\partial^2+\frac{1}{2}c_7\chi^2_v\bigr)
 \end{array} \right) D\bigl(g^\mu_\rho;\rho_{13};x-y\bigr) = \left(
 \begin{array}{cc} 
  g^{\mu \nu} & 0 \\
  0 & 1 
 \end{array} \right)\delta^4\bigl(x-y\bigr),
\end{eqnarray*}
which gives the form in momentum representation neglecting a factor for boundary condition
\begin{eqnarray*}
iD\bigl(g^\mu_\nu\bigr) = i\left[-\frac{1}{k^2-M^2_Y}\left(
 \begin{array}{cc}
  g^\mu_\nu-\frac{k^\mu k_\nu}{M^2_Y} & 0 \\
  0 & 0 
 \end{array} \right)
+ \frac{1}{k^2-\frac {2c_7}{g^2}M^2_Y}\frac{\chi^2_v+\rho^2_v}{\chi^2_v}\left(
 \begin{array}{cc}
  \frac{g^2\rho^2_vk^\mu k_\nu}{4M^4_Y} & -\frac{ig\rho_vk_\nu}{2M^2_Y} \\
 \frac{ig\rho_vk^\mu}{2M^2_Y} & 1
 \end{array} \right)\right].
\end{eqnarray*}
It thus follows that the propagators for the gauge and the scalar fields can not be separated 
for each of the gauge and scalar particles in general in the case of the U gauge.

On the other hand, in the $R_\xi$ gauge the third term in (\ref{3.8}) makes a total derivative 
together with the term from the gauge fixing Lagrangian (\ref{4.3}) to be neglected from it. 
It follows from (\ref{3.9}) and (\ref{4.4}) that the propagators for the gauge particles 
are given in the well known form and thus approach to those for the vector particles in the limit of 
$\xi \rightarrow 0$ for massive particles.

The propagators of the ghost fields are given with the mass term in a form 
$(mass\ of\ vector\ particle)^2/\xi$ except for a mass zero field corresponding to $A_\mu$, 
which remains as a free field, and thus these ghost fields approach to zero in the limit of 
$\xi \rightarrow  0$ in agreement with the known results.

The propagators for the scalar fields are obtained from (\ref{4.1}), (\ref{4.2}) and (\ref{4.4}) and 
some of them are given explicitly as follows
\begin{eqnarray*}
&&
\frac{i}{k^2-\frac{2c_7}{g^2}M^2_Y}\ ;\ \ \ {\rm for}\ \ \ 
 \frac{1}{\sqrt{\rho^2_v+\eta^2_v}}(\rho_v\tilde \eta_{12}-\eta_v \tilde\rho_{11}),\\
&&
\frac{i}{k^2-M^2_Y/{2\xi}}\ ;\ \ \ {\rm for}\ \ \ 
 \frac{1}{\sqrt{\rho^2_v+\eta^2_v}}(\eta_v\tilde \eta_{12}+\rho_v \tilde\rho_{11}),\\
&&
\frac{i}{k^2-\frac{2c_8}{g^2}M^2_X}\ ;\ \ \ {\rm for}\ \ \ 
 \frac{1}{\sqrt{\chi^2_v+\eta^2_v}}(\chi_v\tilde\eta_{13}-\eta_v \tilde\chi_{11}),\\
&&
\frac{i}{k^2-M^2_X/{\xi}}\ ;\ \ \ {\rm for}\ \ \ 
 \frac{1}{\sqrt{\chi^2_v+\eta^2_v}}(\chi_v\tilde\chi_{11}+\eta_v \tilde\eta_{13}),\\
&&
\frac{i}{k^2-M^2_+/\xi}\ ;\ \ \ {\rm for}\ \ \ 
 \tilde\rho_{M_+},\ \ \ \ \ \ 
\frac{i}{k^2-M^2_-/\xi}\ ;\ \ \ {\rm for}\ \ \ 
 \tilde\eta_{M_-}.
\end{eqnarray*} 
It follows that in the limit of $\xi \rightarrow 0$ the mass for the combinations such as 
$(\eta_v\tilde\eta_{12}+\rho_v \tilde\rho_{11})$, $\chi_v\tilde\chi_{11}+\eta_v \tilde\eta_{13}$, 
$\tilde\rho_{M_+}$ and $\tilde\eta_{M_-}$ which are not contained in the potential approaches to 
infinity and thus the fields with the mass will approach to zero, while the fields with the mass 
independent of $\xi$ such as those for $\rho_v\tilde\eta_{12}-\eta_v \tilde\rho_{11}$ and 
$\chi_v\tilde\eta_{13}-\eta_v \tilde\chi_{11}$ can not disappear. 
And it is easily seen that all components of the tilde fields must remain without vanishing. 
Then it is apparent that the condition (\ref{3.3}) of the unitary gauge is not satisfied for such 
$\chi$ with nonzero components $\tilde\chi_{\alpha j}\ (\alpha=1,2;\ j=1,2;\ \alpha =1,\ j=3)$ 
and similarly (\ref{3.6}) is not satisfied for the $\rho$ with the components 
$\tilde\rho_{\alpha j}\ (\alpha =1,2;\ j=1;\ \alpha =1,\ j=2)$. 
Thus, it may be concluded that in general the gauge given by the limit of $\xi \rightarrow 0$ is not 
the unitary gauge given by the condition of the unitary gauge such as (\ref{3.3}) and (\ref{3.6}) 
though the ghost fields and the Goldstone bosons drop out.  
Of course, if only the $\chi$ exists, it is easily seen that the gauge given by the limit of 
$\xi \rightarrow 0$ agrees with the unitary gauge as in the SM because in this case the tilde fields 
except for $\tilde\chi_{23}$ disappear.

It is noted that the eight degree of freedom of the scalar particles disappears in the limit of 
$\xi \rightarrow 0$ which fact agrees with the general result in the Higgs mechanism and then 
the special gauge of $\xi \rightarrow 0$ contains only the physical vector and scalar particles 
in a sense with definite mass, while in the gauge of the Higgs mechanism the vector and 
scalar particles except for $\chi_h,\ \rho_h$ and $\eta_{h21}$ are mixed as seen from (\ref{3.10}) 
to give their propagators and in the sense with no definite mass they will not correspond to 
the physical particles in general.
\begin{acknowledgments}
One of the authors (T.K.) is grateful to Prof T. Maekawa for valuable comments, 
helpful discussions and continuous encouragement.
\end{acknowledgments}
%
%


\begin{thebibliography}{99}
\bibitem{R-gauge}
T.D. Lee and C.N. Yang, Phys. Rev. {\bf 128} (1962), 885.\\
G. 't. Hooft, Nucl. Phys. {\bf B33} (1971), 173.\\
B.W. Lee and J. Zinn-Justin, {\bf D5} (1972) 3121, 3137, 3155; {\bf D7} (1973) 1049.\\
K. Fujikawa, B.W. Lee and A.I. Sanda, Phys. Rev. {\bf D6} (1972), 2923.
\bibitem{FP-ghost}
R.P. Feynman, Acta Phys. Polonica {\bf 26} (1963), 697.\\
B.S. Dewitt, Phys. Rev. {\bf 162} (1967), 1195; 1239.\\
L.D. Fadde'ev and V.N. Popov, Phys. Lett. {\bf 25B} (1967), 29.
\bibitem{SW}
S. Weinberg, Phys. Rev. {\bf D7} (1973), 1068; 2887.
\bibitem{AL}
E.S. Abers and B.W. Lee, Phys. Reports 9 (1973), 1.\\
P. Langacker, Phys. Reports 72 (1981), 185.
\bibitem{PWH}
P.W. Higgs, Phys. Rev. Lett. 12 (1964), 132; 113 (1964), 508.\\
F. Englert and R. Brout, Phys. Rev. Lett. 13 (1964), 321.\\
P.W. Higgs, Phys. Rev. 145 (1966), 1156.\\
T.W. Kibble, Phys. Rev. 155 (1967), 1554.
\bibitem{GOLD}
J. Goldstone, Nuovo Cimento 19 (1961), 15.\\
Y. Nambu and G. Jona-Lasinio, Phys. Rev. 122 (1961), 345; 124 (1961), 246.
\bibitem{TTMH}
T. Kiyan, T. Maekawa, M. Masuda and H. Taira, hep-ph/0206180.
\end{thebibliography}
\end{document}